\documentclass[a4paper]{article}

\usepackage{INTERSPEECH2020}
\usepackage{cite}
\usepackage{bm}
\usepackage{multirow}
\usepackage{xcolor}
\usepackage{mathrsfs}
\usepackage{amsmath}

\title{Self-and-Mixed Attention Decoder with Deep Acoustic Structure for Transformer-based LVCSR}
\name{Xinyuan Zhou$^{1,2}$,Grandee Lee$^2$, Emre Y{\i}lmaz$^2$, Yanhua Long$^{1}$\thanks{This work was done when Xinyuan Zhou was an intern at National University of Singapore. Yanhua Long is the corresponding author. The work is supported by the National Natural Science Foundation of China (No.61701306),  and Human-Robot Interaction Phase 1 (Grant No. 192 25 00054), National Research Foundation (NRF) Singapore under the National Robotics Programme; AI Speech Lab (Award No. AISG-100E-2018-006), NRF Singapore under the AI Singapore Programme; Human Robot Collaborative AI for AME (Grant No. A18A2b0046), NRF Singapore.}, Jiaen Liang$^3$, Haizhou Li$^2$}
\address{
  $^1$Shanghai Normal University, Shanghai, China\\
  $^2$National University of Singapore, Singapore \\
  $^3$Unisound AI Technology Co., Ltd., Beijing, China}
\email{\{xinyuan\_zhou,grandee.lee\}@u.nus.edu, emrey@kth.se, yanhua@shnu.edu.cn, liangjiaen@unisound.com, haizhou.li@nus.edu.sg}

\begin{document}

\maketitle
\begin{abstract}

Transformer has shown impressive performance in automatic speech recognition. It uses an encoder-decoder structure with self-attention to learn the relationship between high-level representation of source inputs and embedding of target outputs. In this paper, we propose a novel decoder structure that features a self-and-mixed attention decoder (SMAD) with a deep acoustic structure (DAS) to improve the acoustic representation of Transformer-based LVCSR. Specifically, we introduce a self-attention mechanism to learn a multi-layer deep acoustic structure for multiple levels of acoustic abstraction. We also design a mixed attention mechanism that learns the alignment between different levels of acoustic abstraction and its corresponding linguistic information simultaneously in a shared embedding space. The ASR experiments on Aishell-1 show that the proposed structure achieves CERs of $4.8\%$ on the dev set and $5.1\%$ on the test set, which are the best reported results on this task to the best of our knowledge.

\end{abstract}

\section{Introduction}
\label{sec:intro}

The sequence-to-sequence (S2S) approach~\cite{sutskever2014sequence} has achieved remarkable results in automatic speech recognition (ASR), in particular, large vocabulary continuous speech recognition (LVCSR)~\cite{chorowski2015attention,bahdanau2016end,chorowski2016towards,zhang2017very,kim2017joint,chan2016listen,huang2019exploring}.
Unlike conventional hybrid ASR, S2S requires neither lexicons, prerequisite models, nor decision trees. It  optimizes the acoustic and language model jointly and simultaneously, learning the mapping directly from speech to text.

The most commonly used structure in S2S approaches is the attention-based encoder-decoder model (AED)~\cite{AED}. This model maps the input feature sequences to the output character sequences and has been widely used in ASR tasks~\cite{chorowski2014end, chan2016listen, Speech-transformer}. Among them, the Listen, Attend and Spell (LAS)~\cite{chiu2018state} structure has shown superior performance to a conventional hybrid system using large amounts of training data. LAS uses an encoder that is a pyramidal recurrent neural network (RNN) to convert low-level speech signals into higher-level acoustic representations, and then the relationship between these representations and targets is learned by an attention mechanism at the RNN-based decoder. However, due to the sequential nature of RNNs, the LAS model doesn't support parallelization of calculations, therefore, is prevented from big data training.

To remedy this problem, new encoder-decoder structures with self-attention networks have recently been proposed~\cite{Transformer}. With self-attention, these structures can not only effectively capture global interaction between sequences~\cite{yang2018modeling}, learning the direct dependence of long-distance sequences~\cite{hochreiter2001gradient}, but also support parallelized model training~\cite{Transformer}. Now, these structures are widely used in a variety of machine learning tasks providing significant improvements. Vaswani et al.~\cite{Transformer} first proposed a S2S self-attention
based model called the Transformer, and it achieves state-of-the-art performance on WMT2014 English-to-French translation task with remarkable lower training cost. For the ASR task, Transformer also uses the AED structure. Unlike LAS, the Transformer uses the multi-head self-attention (MHA) sub-layer to learn the source-target relationship, and capture the mutual information within the sequences to extract the most effective high-level features. This enables Transformer-based ASR systems to achieve competitive performance over the conventional hybrid and other end-to-end approaches~\cite{Speech-transformer,zhou2019improving,pham2019very,li2019speechtransformer,li2019improving,zhou2018syllable,zhou2018comparison}.

Inspired by the Transformer and the layer-wise coordination~\cite{Layer-wise}, we propose a novel decoder structure that features a self-and-mixed attention decoder (SMAD) with a deep acoustic structure (DAS) to improve the acoustic representation of Transformer-based LVCSR. With reference to the standard Speech-Transformer in~\cite{Speech-transformer}, several improvements have been made at the decoder. 

In the Speech-Transformer decoder, the linguistic information is first extracted using a self-attention sub-layer, and then processed together with the encoder output in another source-target attention sub-layer. The same encoder output is repeatedly taken by every decoder layer to establish the acoustic-target relationship. In this paper, 
we propose a new attention block, called self-and-mixed attention (SMA), as an unified attention sub-layer in the decoder, that takes the concatenation of the encoded acoustic representation and the word label embedding as input. In this way, the acoustic and linguistic information is projected into the same subspace in the deep decoder network structure during the attention calculation. 

Furthermore, our decoder learns the acoustic and linguistic information together in a layer-by-layer fashion with the SMA mechanism instead of repeatedly using the same acoustic representations in each decoder layer. 
This is motivated by two intuitions, 1) we hope to benefit from a deep decoder network structure that encodes multi-level of abstraction from both acoustic and linguistic representation, and 2) we hypothesize that a shared acoustic and linguistic embedding space will help the network to learn the association between acoustic and linguistic information, and improve their alignments. 

We will introduce the Transformer-based ASR as the prior work in Section 2, and discuss the details of the new decoder in Section 3. 


\section{Transformer-based ASR}
\label{sec:trans}

\subsection{Encoder-Decoder with Attention}
\label{subsec:ed}

The Transformer model~\cite{Transformer} uses an encoder-decoder structure similar to many other neural sequence transduction models. The encoder can be regarded as a feature extractor, which converts the input vector $\bm{x}$ into a high-level
representation $\bm{h}$. Given $\bm{h}$, the decoder generates prediction sequence $\bm{y}$ one token at a time in an auto-regressive manner. In  an ASR task \cite{Speech-transformer,Transformer-CTC}, tokens are usually modeling units, such as phones, characters or sub-word, etc. 

The encoder has $N$ layers, each of which contains two sub-layers: a multi-head self-attention and a position-wise fully connected feed-forward network (FFN). Similar to the encoder, the decoder is also composed of a stack of $M$ identical layers. In addition to the two sub-layers, each layer of decoder also has a third sub-layer between the FFN and MHA to perform multi-head source-target attention over the output representation of the encoder stack.

\subsection{Multi-Head Attention}
\label{subsec:mhsa}

Multi-head attention is the core module of the Transformer model. Unlike single-head  attention, MHA can learn the relationship between queries, keys and values from different subspaces. It computes the ``Scaled Dot-Product Attention" with the following form:
\begin{equation}
  \text{Attention}(Q,K,V) = \text{softmax}(\frac{QK^T}{\sqrt{d_k}})V,
  \label{eq1}
\end{equation}
where $Q \in \mathbb{R}^{t_{q} \times {d_q}}$ is the query, $K \in \mathbb{R}^{t_{k} \times {d_k}}$ is the key and $V \in \mathbb{R}^{t_{v} \times {d_v}}$ is the value.
$t_*$ are the length of input and $d_*$ are the dimension of corresponding elements. To prevent pushing the softmax into extremely small gradient regions caused by large dimensions,
the $\frac{1}{\sqrt{d_k}}$ is used to scale the dot products.

In order to calculate attention from multiple subspaces, multi-head attention is constructed as follow:
\begin{equation}
\text{MHA}(Q,K,V) = \text{Concat}(Head_1, \cdots, Head_H)W^O, \label{eq2}
\end{equation}
\begin{equation}
 \quad Head_i = \text{Attention}(QW_i^Q, KW_i^K, VW_i^V), \label{eq3}
\end{equation}
where $W_i^*$ is the projection matrix. $W_i^Q \in \mathbb{R}^{d_{\text{model}} \times {d_Q}}$, $W_i^K \in \mathbb{R}^{d_{\text{model}} \times {d_K}}$, $W_i^V \in \mathbb{R}^{d_{\text{model}} \times {d_V}}$ and $W^O \in \mathbb{R}^{d_{\text{model}} \times {d_O}}$, $d_{\text{model}}$ is the dimension of the input vector to the encoder, $H$ is the number of heads.
For each $Q$, $K$, $V$ in each attention, they are projected to $d_*$ dimensions through three linear projection layers $W_i^*$ respectively. After performing $H$ attentions, the outputs are then concatenated and projected again to obtain the final values.

\subsection{Positional Encoding}
\label{subsec:pe}

Unlike RNN, the MHA contains no recurrence and convolution, it cannot model the order of the input acoustic sequence. 
We follow the idea of ``positional encoding" that is added to the input as described in \cite{Transformer}.

\begin{figure}[t]
  \centering
  \includegraphics[width=7cm]{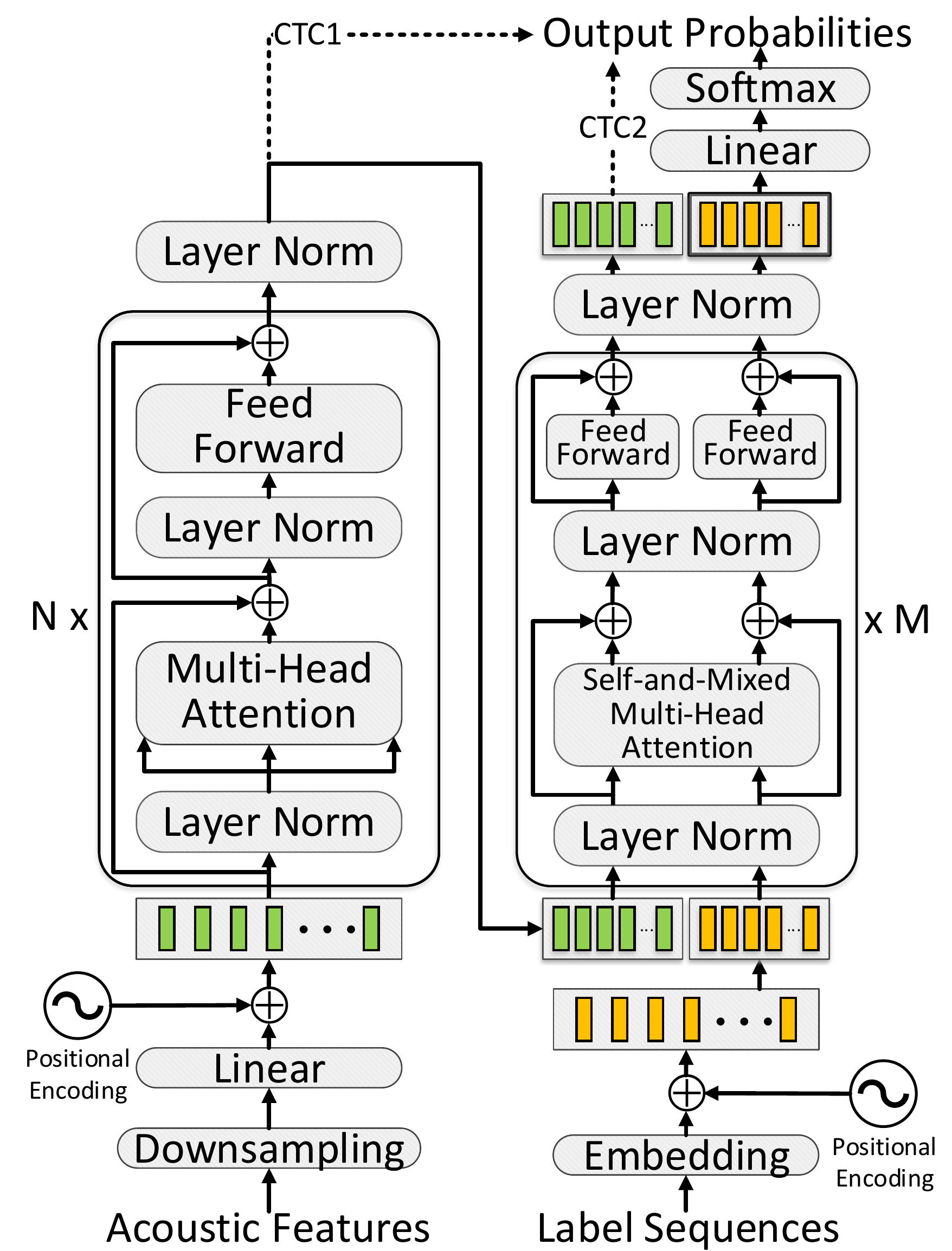}
  \caption{The system architecture of the Transformer-based ASR with the proposed self-and-mixed attention decoder.}
  \label{fig:bert}
\end{figure}

\section{Self-and-Mixed Attention Decoder}

\subsection{Architecture}
\label{subsec:arch}

Figure~\ref{fig:bert} shows an encoder-decoder architecture for ASR. We adopt the same encoder as in Transformer, and propose a Self-and-Mixed Attention Decoder (SMAD), that is an attention-based auto-regressive structure. The main difference between SMAD and the standard Transformer decoder\cite{Speech-transformer} lies in the ways we integrate acoustic representations, $\bm{h}$.

Firstly, unlike the decoder in Transformer which takes the same $\bm{h}$ repeatedly to every decoder layer, SMAD employs a deep acoustic structure (DAS), which is a $M$-layer network to capture multiple level of acoustic abstraction. For simplicity, we use a single-head self-and-mixed attention in Figure \ref{fig:Mix-A} to illustrate the Self-and-Mixed MHA component in the decoder layer of Figure \ref{fig:bert}, where the self-attention handles the acoustic representations, and the mixed-attention handles the acoustic-target alignment.  With $M$ decoder layers stacking in a serial pipeline, the flow of acoustic information in Figure \ref{fig:Mix-A} (green) forms a deep acoustic structure.


Secondly, the decoder in the standard Transformer uses a self-attention module to learn the current target representation based on the previous tokens and learn the acoustic-targets dependencies using another separate source-target attention sub-layer. However, in SMAD, we merge these two attentions into one as illustrated in Figure \ref{fig:Mix-A}. We concatenate the encoded acoustic representation and linguistic targets to form a joint embedding as the input to the decoder layer. After the self-and-mixed MHA, the concatenated representation with both acoustic and linguistic information is fed to the FFN and next self-and-mixed MHA. Since the information flow in the proposed decoder contains two modalities, we also employ modality-specific residual connections and  position-wise feed forward networks to separate linguistic and acoustic information before obtaining the posterior probability at the output of the softmax layer preceded by a linear layer. 

We perform the same downsampling as in \cite{Transformer-CTC} before the encoder using two $3\times3$ CNN layers with stride 2 to reduce the GPU memory occupation and and the length of the input sequence.

\subsection{Self-and-Mixed Attention (SMA)}
\label{sec:self-mixed-attention}
\begin{figure}[t]
  \centering
  \includegraphics[width=5cm]{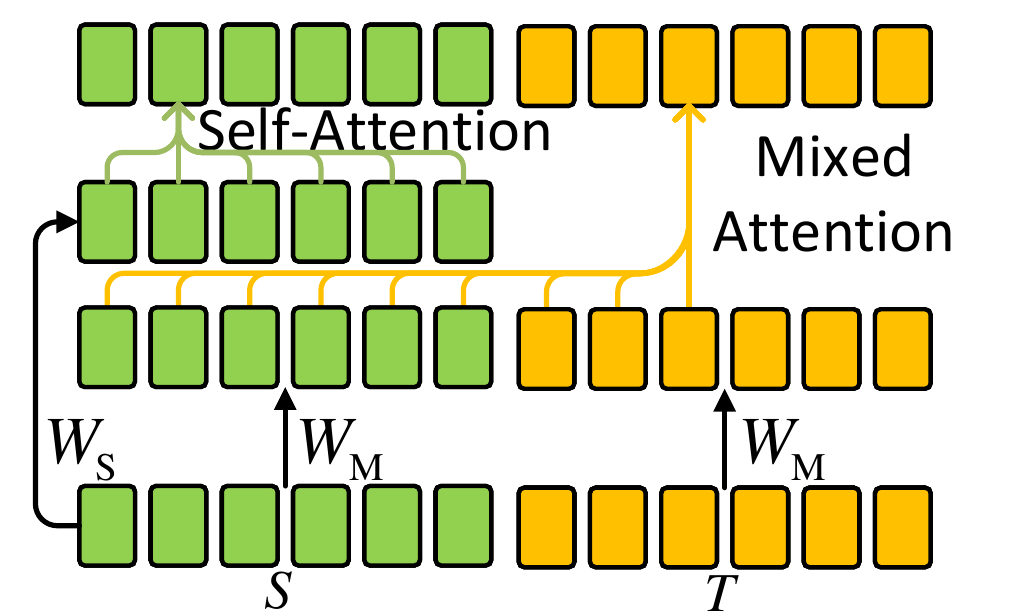}
  \caption{A single-head  self-and-mixed attention mechanism as a sub-layer of the decoder in Figure \ref{fig:bert}.}
  \label{fig:Mix-A}
\end{figure}

For simplicity, we take one head of the self-and-mixed MHA in Figure~\ref{fig:bert} as an example.
The SMA consists of two independent attention mechanisms: a self-attention 
for acoustic-only representation, and a mixed attention to learn linguistic representation and the acoustic-target association. As shown in Figure~\ref{fig:Mix-A}, $S$ refers to source, which is the acoustic representation (marked in green) and $T$ refers to target, which is the linguistic information (marked in yellow).

Specifically, for self-attention in the SMA, $Q$, $K$, $V\in \mathbb{R}^{n \times {d_\text{model}}}$ are projected by $W^Q_{\text{S}}$, $W^K_{\text{S}}$, $W^V_{\text{S}}$ from $S$ respectively, with $n$-length acoustic representation. The acoustic hidden representation in the current layer is generated using the accumulated acoustic information in the previous layer using the self-attention mechanism.

For the mixed attention, a linguistic token in the current layer is generated using the acoustic hidden representation and the preceding linguistic tokens in the previous layer using a mixed attention mechanism.
The mixed attention is formulated as follows:
\begin{align}
\text{MixedAtt}(Q,K,V) &= \text{softmax}(\frac{QK^T}{\sqrt{d_{\text{model}}}}+Mask)V,\label{eq6} \\
    Q &= TW^Q_{\text{M}}, \\
    K &= \text{Concat}(S,T)W^K_{\text{M}}, \\
    V &= \text{Concat}(S,T)W^V_{\text{M}}, \\
Mask(i,j) &= \left\{
\begin{aligned}
-\infty & , & j>i+n \\
0 & , & \text{otherwise}
\end{aligned}
\right.,
\end{align}
where $Q \in \mathbb{R}^{m \times {d_\text{model}}}$, $K$, $V\in \mathbb{R}^{(n+m) \times {d_\text{model}}}$, and $Mask \in \mathbb{R}^{m \times (n+m)}$ is the mask matrix, $m$ is the number of tokens, $i$ and $j$ refer to the index of row and column of $Mask$.
To project the acoustic and linguistic information into the same subspace,  we concatenate $S$ and $T$ and apply the same projection matrix $W_{\text{M}}$ for $K$ and $V$. We use the acoustic representation $\bm{h}$ of entire sentence for the decoding of tokens, at the same time, we introduce a 
$Mask$ to ensure that the prediction of token sequence is causal, i.e., when predicting a token, we only use information of the tokens before it.  When $Mask(i,j)$ equals to $-\infty$, the corresponding position in softmax output will approach zero, which prevents position $i$ from attending to position $j$.

\subsection{Multi-Ojective Learning}

As often done in encoder-decoder structures, our model also uses the connectionist temporal classification (CTC)~\cite{watanabe2017hybrid,graves2006connectionist,Transformer-CTC} to benefit from the monotonic alignment. The CTC loss function~\cite{watanabe2017hybrid} is used to jointly train the extraction of acoustic representation by  Multi-Objective Learning~(MOL):
\begin{equation}
\mathcal{L}_{\text{MOL}} = \lambda \text{log}P_{\text{ctc}}(C|X) + (1-\lambda) \text{log}P^*_{\text{att}}(C|X), \label{eq10}
\end{equation}
\begin{align}
P^*_{\text{att}}(C|X) &= \prod_{l=1}^LP(c_{l}|c^*_{1},\cdots,c^*_{l-1},X), \label{eq11} \\
P_{\text{ctc}}(C|X) &= \sum_{Z}\prod_{t=1}^TP(z_{t}|z_{t-1},C)P(z_{t}|X), \label{eq12} \\
P(z_{t}|X)&=\text{Softmax}(\text{LinB}(\textbf{h}_t)),     \label{eq13}
\end{align}
where $\mathcal{L}_{\text{MOL}}$ is the multi-objective loss with a tuning parameter $\lambda\in[0,1]$, $P^*_{\text{att}}(C|X)$ is the Transformer loss modeled by Kullback-Leibler divergence~\cite{szegedy2016rethinking} loss .
$X=\{x_{t}\in\mathbb{R}^{D}|t=1,\cdots,T\}$ is a $T$-length speech feature sequence,
$x_{t}$ is a $D$-dimensional speech feature vector at frame t, $C=\{c_t \in \mathcal{U}|l = 1,\cdots,L\}$ is an $L$-length letter sequence containing all the characters $\mathcal{U}$ in this task, and $c^*_{l-1}$ is the ground truth of $c^*_{l}$'s previous token. $Z=\{z_{t}\in\mathcal{U}\cup{\langle \text{b}\rangle}|t=1,\cdots,T\}$
is a framewise letter sequence with an additional blank symbol ``$\langle b\rangle$", and $\textbf{h}_t$ is the acoustic hidden representation vector, $\text{LinB}(\cdot)$is a linear layer to convert $\textbf{h}_t$ to a $(|\mathcal{U}|+1)$ dimensional vector.

We explore two different locations where the CTC loss can be applied as shown in Figure \ref{fig:bert}. For CTC1,
\begin{equation}
\textbf{h}_t=\text{Encoder}_{t}(X),
\end{equation}
$\text{Encoder}_t(\cdot)$ accepts the full feature sequence $X$ and output acoustic representation $\textbf{h}_t$ at t. Similar to the previous techniques, this CTC loss is used to jointly train the encoder.

For CTC2, since the acoustic representation is also updated in the SMAD, $\textbf{h}_t$ is produced as follows:
\begin{equation}
\textbf{h}_t=\text{Decoder}^A_t(\text{Encoder}(X)),
\end{equation}
where the $\text{Decoder}^A_t(\cdot)$ is the acoustic side of decoder stack output at t. In this way, the entire acoustic representation extraction process is jointly trained using the CTC loss.

\section{Experiments}
\subsection{Experimental setup}
We conduct experiments on 170 hours Aishell-1~\cite{aishell_2017} using the ESPnet~\cite{watanabe2018espnet} end-to-end speech processing toolkit. For all experiments, we extract 80-dimensional log Mel-filter bank plus pitch and its $\Delta$, $\Delta\Delta$ as acoustic features and normalize them with global mean computed from the training set. The frame-length is 25 ms with a 10 ms shift.

The standard configuration of the state-of-the-art ESPnet Transformer recipe on Aishell-1 is used for both the baseline and proposed model. Each model contains 12-layer encoder and 6-layer decoder, where the $d_{\text{model}}=256$ and the dimensionality of inner-layer in FFN $d_{\text{ff}}=2,048$. In all attention sub-layers, 4 heads are used for MHA. The whole network is trained for 50 epochs and warmup~\cite{Transformer} is used for the first 25,000 iterations. We use 4,230 Chinese characters which are extracted from the train 
set as modeling unit. A ten-hypotheses-width beam search is used with the the one-pass decoding for CTC as described in \cite{watanabe2017hybrid} and a two-layer RNN language model (LM) shallow fusion~\cite{RNN-LM,shallowfusion},
which was trained on the training transcriptions of Aishell-1 with 4,230 Chinese characters. We also evaluate the effect of speed perturbation (SP)~\cite{ko2015audio}, SpecAugment (SpecA)~\cite{park2019specaugment} and CTC joint training in our experiments.

\subsection{Results and Discussion}
\label{subsecrst}

Table~\ref{tab:Aishell} reports the results of the proposed Transformer-based ASR, referred to as T-SMAD, the conventional Kaldi hybrid~\cite{kaldi} and other E2E ASR systems. Shallow fusion with 5-gram language model is used in both \cite{LAS_Aishell,Transducers_Aishell}. In ESPnet RNN, Transformer\cite{ESPnet_result} and T-SMAD, the RNN LM was also used for shallow fusion. We consider the Transformer with speed perturbation and CTC in ESPnet as our reference baseline.

According to Table 1, the T-SMAD system with the proposed CTC2 outperforms all other systems, including both the Transformer baseline and the Kaldi hybrid systems. A relative 20.0\% CER reduction on the test set is obtained over the best hybrid system (chain). A relative 10\% CER reduction on the dev set and 10.4\% CER reduction on the test set is reported over the best E2E system (baseline). Moreover, it can be seen that CTC2 provides better ASR results than CTC1 due to the fact that the acoustic feature extraction of the entire network and the decoder are jointly trained in CTC2. In these experiments, the default parameter $\lambda=0.3$, which is tuned for the baseline system in ESPnet, is used for both CTC1 and CTC2.  

\begin{table}[t]
  \caption{Results comparsion on Aishell-1 in CER\%}
  \label{tab:Aishell}
  \centering
  \begin{tabular}{l l r r}
    \toprule
    &System                  &Dev            &Test  \\
    \midrule
    &Kaldi (chain)          & -          & 7.5             \\
    &Kaldi (nnet3)          & -          & 8.6              \\
    &LAS\cite{LAS_Aishell}                      & -          & 10.6              \\
    &ESPnet RNN\cite{ESPnet_result}           & 6.8          & 8.0              \\
    &RNN-T\cite{Transducers_Aishell}          & 10.1          & 11.8  \\
    &Transformer +SP+CTC (baseline)\cite{ESPnet_result}            &6.0          &6.7                \\
    \midrule
    &T-SMAD +SP+CTC1              &5.9          &6.4                \\
    &T-SMAD +SP+CTC2              &5.4          &6.0                \\
    \bottomrule
  \end{tabular}
\end{table}


In addition to the default configuration of ESPnet, we further implement the SpecAugment in our system to investigate its impact on the ASR performance, all experiments are with RNN LM. As shown in Table~\ref{tab:Aishell-SpecA}, the baseline system with SpecAugment gives a relative CER reduction of 13.3\% on dev set and 17.9\% on test set. T-SMAD with SpecAugment continues to outperform the corresponding Transformer baseline with SpecAugment by relative 8.6\% on dev and 9.4\% on the test set. The best performing system, T-SMAD+SP+SpecA+CTC2 achieves a CER of 4.8\% and 5.1\% on the dev and test set, respectively. To the best of authors' knowledge, these are the best results reported on the Aishell-1 corpus. It can be concluded that the proposed SMAD achieves improved alignment due to the deep acoustic structure and the mixed attention, and yields consistent performance improvements over the standard Transformer architecture.

\begin{table}[t]
  \caption{Results (CER\%)  with SpecAugment on Aishell-1}
  \label{tab:Aishell-SpecA}
  \centering
  \begin{tabular}{l l l l r r}
    \toprule
    &System                         &&&Dev            &Test  \\
    \midrule
    &Transformer +SP+SpecA           &&&5.8          &6.4                \\
    &Transformer +SP+SpecA+CTC       &&&5.2          &5.5               \\
    \midrule
    &T-SMAD +SP+SpecA             &&&5.3          &5.8                \\
    &T-SMAD +SP+SpecA+CTC1        &&&5.0          &5.4                \\
    &T-SMAD +SP+SpecA+CTC2        &&&4.8          & 5.1               \\
    \bottomrule
  \end{tabular}
\end{table}

To examine the contribution of each component in SMAD, we perform several ASR experiments by removing the encoder, DAS, mixed attention and modelity-specific network one at a time.
To focus on the SMAD mechanism, all the results are produced without additional LM and are reported in Table~\ref{tab:Ablation}.

Firstly, we remove the encoder in T-SMAD. For a fair comparison,
we increase the number of decoder layers to 18 in order to keep the number of model parameters the same. Directly concatenating the acoustic features with linguistic targets as the input to the decoder increases the CER from 6.1\% to 8.2\% on the test set, that suggests the encoder block is essential for effective acoustic representation. Secondly, we give the same encoder acoustic representation to each SMAD layer in the say way as the standard Transformer, without the deep acoustic structure (`T-SMAD w/o DAS'). This system gives a higher CER than T-SMAD, that confirms the effectiveness of the deep acoustic structure. Thirdly, we replace the mixed attention with the two standard attention mechanisms of Transformer to extract the linguistic features and learn the source-target alignment, respectively. We observe that the removal of mixed attention degrades the performance, that suggests that mapping acoustic-linguistic into the same subspace does help to learn a better alignment. Lastly, the contribution of a modality-specific network has been found to be less prominent than the previous components. It is worth noting that even without modality-specific network, T-SMAD still outperforms the standard Transformer without any increase in the number of model parameters.

\begin{table}[t]
  \caption{Contribution of each component in SMAD architecture to the ASR performance on Aishell-1. 
  All experiments are with speed perturbation and SpecAugment.}
  \label{tab:Ablation}
  \centering
  \begin{tabular}{l l r r}
    \toprule
    &System                             &Dev        &Test  \\
    \midrule
    &T-SMAD                          &5.6          &6.1             \\
    &Transformer                     &6.7           &7.4           \\
    \midrule
    &T-SMAD w/o encoder              &7.5         &8.2              \\
    &T-SMAD w/o DAS     &6.6          &7.3                \\
    &T-SMAD w/o mixed attention      &6.2          &6.9                \\
    &T-SMAD w/o modality-specific network &5.8          &6.3                \\
    \bottomrule
  \end{tabular}
\end{table}

\section{Conclusion}

We propose a novel decoder structure for Transformer-based LVCSR that features a self-and-mixed attention decoder (SMAD) with a deep acoustic structure (DAS) to improve the acoustic representation. With SMAD mechanism, we have studied the interaction between acoustic and linguistic representation in the training and decoding of LVCSR system, that opens up a promising future direction for improving E2E ASR systems. We confirm  that SMAD and DAS effectively improve the acoustic-linguistic representation in the decoder. The performance gain is attributed to the self-and-mixed attention mechanism that improves the acoustic-linguistic association and alignment in the Transformer decoder.
The proposed technique has achieved the best results ever reported on both the dev and test sets of Aishell-1. Furthermore, we also investigate the impact of the components of the SMAD on the ASR performance and validate their effectiveness.

\bibliographystyle{IEEEtran}

\bibliography{mybib}

\end{document}